\documentclass[journal]{IEEEtran}
\usepackage{mathrsfs}
\usepackage{tabu}
\usepackage{bbm}
\usepackage{amsmath,amscd,amssymb,latexsym}
\usepackage[all]{xy}
\usepackage{mathrsfs}
\usepackage{pgf,tikz}
\usepackage{tabu}
\usepackage{graphicx}
\usepackage{multicol}

\newtheorem{theorem}{Theorem}[section]
\newtheorem{lemma}[theorem]{Lemma}

\newtheorem{definition}[theorem]{Definition}
\newtheorem{example}[theorem]{Example}

\newtheorem{remark}[theorem]{Remark}

\newlength{\halfpagewidth}
\divide\halfpagewidth by 2

\ifCLASSINFOpdf
\else
\fi
\begin{document}
%
\title{ Twisted Reed-Solomon Codes With One-dimensional Hull }
%
%
%

\author{Yansheng Wu~
 \thanks{Manuscript received August 31, 2020; accepted September  16, 2020. This research was supported by the National Natural Science Foundation of China (No.
61772015). The associate editor coordinating the review of this letter and approving it for publication was  Yu-Chih Huang. }
\thanks{Yansheng Wu is with the  School of Computer Science, Nanjing University of Posts and Telecommunications, Nanjing, 210023, P. R. China (e-mail: wysasd@163.com).}

\thanks{Digital Object Identifier ~~~~~~~~~}
}

\maketitle

\begin{abstract} The hull of a linear code is defined to be the intersection
of the code and its dual.  When the size of the hull is small, 
it has been proved that some algorithms for checking permutation equivalence
of two linear codes and computing the automorphism
group of a linear code are very effective in general.  Maximum distance separable (MDS) codes are codes meeting the Singleton bound. Twisted Reed-Solomon codes is a  generalization of   Reed-Solomon codes, which is also a nice construction for MDS codes. In this short letter, we obtain some  twisted Reed-Solomon  MDS codes with one-dimensional hull. Moreover, these codes  are not monomially equivalent to  Reed-Solomon codes.


\end{abstract}

\begin{IEEEkeywords}
twisted Reed-Solomon codes, one-dimensional hull, monomially equivalent.
\end{IEEEkeywords}

%
\IEEEpeerreviewmaketitle

\section{Introduction}
%
%
%
%

\IEEEPARstart{G}{iven}  a linear code $\mathcal C$ of length $n$ over the finite field $\Bbb F_q$,
the  dual code of $\mathcal C$  is defined by
$$\mathcal C^{\bot} = \{ {\bf x}\in \Bbb F_q^{n} \mid {\bf x}{\bf y} ^T=0\mbox{ for  all } {\bf y}\in \mathcal C\}$$
where ${\bf x}{\bf y} ^T$ denotes the standard inner product of two vectors ${\bf x}$ and ${\bf y}$. 
The hull of the linear code $\mathcal{C}$ is defined to be
$$\mbox{Hull}(\mathcal{C}) := \mathcal{C} \cap  \mathcal{C}^{\bot}.$$
It is clear that  $\mbox{Hull}(\mathcal{C})$ is also a linear code over $\Bbb F_q$. The hull was originally introduced in 1990 by Assmus, Jr. and Key \cite{AK} to classify finite projective planes.
It had been
shown that the hull plays an important role in determining
the complexity of algorithms for checking permutation equivalence
of two linear codes and computing the automorphism
group of a linear code (see \cite{L1,L2}, \cite{S1}-\cite{SS}), which are very effective in general when the dimension of the hull is small.

It is worth mentioning that the special case of the hulls of linear codes is of much interest. Namely the codes with trivial intersection with its dual, which is also named linear complementary dual (LCD) codes. Massey \cite{M} first introduced this class of codes and proved that there exist asymptotically good LCD codes. A practical application of binary LCD codes against side-channel attacks (SCAs) and fault injection attacks (FIAs) was investigated by Carlet et al. \cite{BCCGM} and Carlet and Guilley \cite{CG}. Since then, the study of LCD codes is thus becoming a hot research topic in coding theory (\cite{CMTQP}-\cite{J}, \cite{L,LDL,LLDL,LCC},\cite{WY}-\cite{WYZY}).
Some nice progress on linear codes with small hulls has been made, for examples (\cite{CLM, LZ}).

A {\em maximum distance separable} (MDS) code has the greatest error correcting capability
when its length and dimension is fixed. MDS codes are extensively used in communications (for example,
 Reed-Solomon codes are all MDS codes), and they have good applications in minimum storage codes and quantum codes.  There are many known constructions for MDS codes; for instance, {\em Generalized Reed-Solomon} (GRS) codes \cite{RS}, based on the equivalent problem of finding $n$-arcs in projective geometry \cite{MS}, circulant matrices \cite{RL}, Hankel matrices \cite{RS2}, or extending GRS codes.

Recently the authors in \cite{FFLZ,LCC}  investigated the hull of MDS codes via generalized Reed-Solomon codes over finite fields. Beelen {\em et al.} \cite{BPR} first gave the definition of  twisted Reed-Solomon codes, which is a generalization of the Reed-Solomon codes, and they proved under some conditions twisted Reed-Solomon codes could be not monomially equivalent to the Reed-Solomon codes.  However, the hull of twisted Reed-Solomon codes have not been studied in that paper. Recently, Wu, Hyun and Lee \cite{WHL} constructed some LCD twisted Reed-Solomon codes.

 
  In this letter, as a follow-up work we will focus on the hull of twisted Reed-Solomon codes. In particular, we will consider to construct some  twisted Reed-Solomon MDS codes with one-dimensional hull, which are not monomially equivalent to Reed-Solomon codes. 
The rest of this letter is organized as follows. In Section II, we introduce basic concepts on the hull of linear codes and twisted Reed-Solomon codes. In Sections III, we present our main results and give some examples. 
We  conclude the letter in Section IV.

\section{Preliminaries}
Let $\Bbb F_q$ be the finite field of order $q$, where $q$ is a prime power.
An $[n, k]_q$ linear code $\mathcal{C}$ over $\Bbb F_q$  is a $k$-dimensional
subspace of $\Bbb F_q^n$. The minimum distance d of
a linear code $\mathcal{C}$ is bounded by the so-called {\em Singleton bound }: $d\le n-k+1$.
If $d= n-k+1$, then the code $\mathcal{C}$ is called a {\em maximum distance separable} (MDS) code.

The following lemma on the hull of linear codes, which is very important for obtaining our main results.

\begin{lemma}\cite[Proposition 1]{LZ} Let $\mathcal C$ be an $[n, k]$ linear code over $\Bbb F_q$ with
generator matrix $G$. Then the code $\mathcal C$ has one-dimensional hull if and only if the rank of the matrix $GG^T$ is  $k-1$, where $G^T$ denotes the transpose of $G$.
\end{lemma}

Recall that a \emph{monomial matrix} is a square matrix which has exactly one nonzero entry in
each row and each column. 

\begin{definition}{\rm Let $\mathcal C_1$ and $\mathcal C_2$ be two linear codes  of the same length over $\Bbb F_q$,  and let $G_1$ be a generator matrix of $\mathcal C_1$. Then
 $\mathcal C_1$ and $\mathcal C_2$ are {\em monomially equivalent} if
 there is a monomial matrix $M$ such that $G_1M$ is a generator matrix of $\mathcal C_2$.
}
 \end{definition}

Next we will recall some constructions of MDS codes.
We begin with the well-known generalized Reed-Solomon codes.

 \begin{definition}{\rm \label{def1} Let $\alpha_{1},\ldots,\alpha_{n}$ be distinct elements in $\Bbb F_{{q}} \cup \{\infty\}$  and $v_{1},\ldots,v_{n}$ be nonzero elements in $\Bbb F_{{q}}$. For $1\leq k \leq n$, the corresponding {\em generalized Reed-Solomon $(GRS)$ code} over $\Bbb F_{{q}}$ is defined by
$$\resizebox{9cm}{!}{ $GR{S_k}({\boldsymbol{ \alpha}},{\bf v}): = \left\{({v_1}f({\alpha_1}), \ldots ,{v_{n }}f({\alpha_{n }})) \mid f(x) \in {\Bbb F_{q}}[x],\phantom{.} \deg(f(x)) < k \right\},$}$$ where $\boldsymbol{ \alpha}=(\alpha_{1},\alpha_{2},\ldots,\alpha_{n})\in (\Bbb F_{{q}}\cup \{\infty\})^{n}$ and ${\bf v}=(v_{1},v_{2},\ldots,v_{n})$, and the quantity $f(\infty)$ is defined as the coefficient of $x^{k-1}$ in the polynomial $f$. 
}
 \end{definition}

If $v_i=1$ for every $i=1, \ldots, n$, then $GR{S_k}({\boldsymbol{ \alpha}},{\bf v})$ is called a {\it Reed-Solomon code.} In fact,   $GRS_k(\boldsymbol{\alpha},{\bf v})$ has a generator matrix  as follows: $$\resizebox{9cm}{!}{ $\left(\begin{array}{cclc} v_1 &v_2 &\ldots  &v_n\\ v_1\alpha_1 &v_2\alpha_2 &\ldots & v_n\alpha_{n}\\ \vdots &\vdots &\ddots&\vdots \\ v_1\alpha_1^{k-1} &v_2\alpha_2^{k-1} &\ldots & v_n\alpha^{k-1}_{n}\end{array}\right)=\left(\begin{array}{cclc} 1 &1 &\ldots  &1\\ \alpha_1 &\alpha_2 &\ldots &\alpha_{n}\\ \vdots &\vdots &\ddots&\vdots \\ \alpha_1^{k-1} &\alpha_2^{k-1} &\ldots & \alpha^{k-1}_{n}\end{array}\right) \left(\begin{array}{cclc} v_1 &0 &\ldots  &0\\ 0 &v_2 &\ldots &0\\ \vdots &\vdots &\ddots&\vdots \\ 0 &0 &\ldots & v_{n}\end{array}\right).$}$$
 It is well-known that a generalized Reed-Solomon code $GRS_k(\boldsymbol{\alpha},{\bf v})$ is an  $ {\left[ {n,k,n-k+1} \right] }$ MDS code and it is monomially equivalent to a Reed-Solomon code.

In 2007, Beelen {\em et al.} \cite{BPR} presented a generalization of  Reed-Solomon codes, so-called {\it twisted Reed-Solomon codes}.

 \begin{definition} \label{def2} Let $\eta$ be a nonzero element in the finite field $\Bbb F_q$. Let $k,t$ and $h$ be nonnegative integers such that $0\le h<k\le q$, $k<n$, and $0<t\le n-k$.  Let $\alpha_{1},\ldots,\alpha_{n}$  be distinct elements in $\Bbb F_{{q}} \cup \{\infty\}$, and we write $\boldsymbol{\alpha}=(\alpha_{1},\alpha_{2},\ldots,\alpha_{n})$. Then the corresponding {\em twisted Reed-Solomon code} over $\Bbb F_q$ of length $n$ and dimension $k$ is given by
\begin{eqnarray*}\resizebox{9cm}{!}{ $\mathcal C_k(\boldsymbol{\alpha}, t,h,\eta)=
\{(f(\alpha_1),  \cdots, f(\alpha_{n})): f(x)=\sum_{i=0}^{k-1}a_ix^i+\eta a_hx^{k-1+t}\in \Bbb F_q[x]\}$}.\end{eqnarray*}

 \end{definition}

 In fact,  \begin{equation}\resizebox{8cm}{!}{ $G=\left(\begin{array}{cclcc}
1 &1 &\ldots  &1\\
 \alpha_1 &\alpha_2 &\ldots & \alpha_{n}\\
   \vdots &\vdots &\ddots&\vdots \\
   \alpha_1^{h-1} &\alpha_2^{h-1} &\ldots & \alpha^{h-1}_{n}\\
   \alpha_1^h+\eta \alpha_1^{k-1+t} &\alpha_2^h+\eta \alpha_2^{k-1+t} &\ldots & \alpha_{n}^h+\eta \alpha_n^{k-1+t}\\
  \alpha_1^{h+1} &\alpha_2^{h+1} &\ldots & \alpha^{h+1}_{n}\\
  \vdots &\vdots &\ddots&\vdots \\
  \alpha_1^{k-1} &\alpha_2^{k-1} &\ldots & \alpha^{k-1}_{n}\end{array}\right)$  } \end{equation} is the generator matrix of the twisted Reed-Solomon code $\mathcal C_k(\boldsymbol{\alpha}, t,h,\eta)$.


Note that in general, the twisted Reed-Solomon codes are not MDS. Beelen {\em et al.} got some results on the twisted Reed-Solomon codes as follows:
 \begin{lemma}\label{lem2} \cite[Theorem 17]{BPR} {\rm  Let $\Bbb F_s \subset \Bbb F_q$ be a proper subfield and $\alpha_{1},\ldots,\alpha_{n}\in \Bbb F_s$. If $\eta \in\Bbb F_q \backslash \Bbb F_s $, then the twisted Reed-Solomon code $\mathcal C_k(\boldsymbol{\alpha}, t,h,\eta)$ is MDS.}
 \end{lemma}

  \begin{lemma} \cite[Theorem 18]{BPR} {\rm  Let  $\alpha_{1},\ldots,\alpha_{n}\in \Bbb F_q$ and $2<k<n-2$. Furthermore, let $H\subseteq \Bbb F_q$ satisfy that the  twisted Reed-Solomon code $\mathcal C_k(\boldsymbol{\alpha}, t,h,\eta)$ is MDS for every $\eta\in H$. Then there are at most $6$ choices of $\eta\in H$ such that $\mathcal C_k(\boldsymbol{\alpha}, t,h,\eta)$ is monomially equivalent to a Reed-Solomon code.

   }
 \end{lemma}

  \begin{lemma} \label{lem3} \cite[Corollary 20]{BPR} {\rm Let $\Bbb F_s \subset \Bbb F_q$ with $|\Bbb F_q \backslash \Bbb F_s|>6$. Let $2<k<n-2$ and $n\le s$. Then there exists $\eta\in \Bbb F_q \backslash \Bbb F_s$ such that $\mathcal C_k(\boldsymbol{\alpha}, t,h,\eta)$ is MDS but not monomially equivalent to a  Reed-Solomon code. }
 \end{lemma}

Throughout the paper, if a code is not monomially equivalent to a Reed-Solomon code, then we call it a code of {\em non-Reed-Solomon type} or a {\em  non-Reed-Solomon code}.

\section{ Twisted Reed-Solomon codes with one-dimensional hull}

Let $\gamma$ be a primitive element of $\Bbb F_q$ and $k\mid (q-1)$. Then $\gamma^{\frac{q-1}{k}}$ generates a subgroup of $\Bbb F_q^*$ of order $k$. Let $\alpha_i=\gamma^{\frac{q-1}{k}i}$ for $1\le i\le k$. One can easily check that  \begin{equation}\label{eq3}
\theta_f=\alpha_1^f+\cdots+\alpha_{k}^f=\left\{
\begin{array}{ll}
  k   &      \mbox{if}\ f\equiv0\pmod {k},\\
0 & \mbox{otherwise}.
\end{array} \right.
\end{equation}


\begin{lemma}\label{lem7}{\rm  Let $q$ be a power of two. If $k$ is a positive integer with $k\mid (q-1)$, $k<(q-1)$, and $h>0$, then there exists a $[2k,k]_q$  twisted Reed-Solomon code $\mathcal C_k(\boldsymbol{\alpha}, t,h,\eta)$ over $\Bbb F_q$ with one-dimensional hull   for
$\boldsymbol{\alpha}=(\alpha_1,\ldots,\alpha_k,\gamma\alpha_1,\ldots,\gamma\alpha_k)$,
where $\gamma$ is a primitive element of $\Bbb F_q$ and $\alpha_i=\gamma^{\frac{q-1}{k}i}$ for $1\le i\le k$.

}


\end{lemma}

{\bf Proof} By Definition \ref{def2}, to make sure that $\mathcal C_k(\boldsymbol{\alpha}, t,h,\eta)$ is a twisted Reed-Solomon code, we need $k\neq q-1$.
From  (1), we recall that $G$ is a generator matrix of  the twisted Reed-Solomon code $\mathcal C_k(\boldsymbol{\alpha}, t,h,\eta)$ over $\Bbb F_q$.


Let
\begin{eqnarray*}\resizebox{8.5cm}{!}{$A_{\beta}=\left( \begin{array}{cccccc}
1 & 1& \ldots  & 1 &1\\
\beta\alpha_1& \beta\alpha_2& \ldots   &  \beta\alpha_{k-1}  & \beta\alpha_k \\
\vdots& \vdots& \ldots  & \vdots &\vdots\\
(\beta\alpha_1)^{k-1}& (\beta\alpha_2)^{k-1} &\cdots  & (\beta\alpha_{k-1})^{k-1} &(\beta\alpha_k)^{k-1} \\
\end{array} \right).$}
\end{eqnarray*}
By  (\ref{eq3}), we have
\begin{eqnarray*} \label{eq5}A_{\beta}A_{\beta}^T=\left( \begin{array}{ccccccc}
k &0&0& \ldots  & 0 &0\\
0& 0&0& \ldots   & 0  & \beta^k k \\
\vdots&\vdots& \vdots& \ldots  & \vdots &\vdots\\
0&0&\beta^k k&\cdots  & 0&0 \\
0&\beta^k k&0&\cdots  & 0&0 \\
\end{array} \right).
\end{eqnarray*}
Let $C_{\beta}=A_{\beta}+B_{\beta}$, where
\begin{eqnarray}\resizebox{8.9cm}{!}{$
B_{\beta}=\left( \begin{array}{cccccc}
0 & 0& \ldots  & 0 &0\\
\vdots& \vdots& \ldots  & \vdots &\vdots\\
\eta(\beta\alpha_1)^{k-1+t}& \eta(\beta\alpha_2)^{k-1+t}& \ldots   &  \eta(\beta\alpha_{k-1})^{k-1+t}  &\eta(\beta\alpha_k)^{k-1+t} \\
\vdots& \vdots& \ldots  & \vdots &\vdots\\
0 & 0& \ldots  & 0 &0\\
\end{array} \right)\begin{array}{l} \\   \\\leftarrow(h+1)th\\  \\ \\  \end{array}$}.\end{eqnarray}
Let $\theta_j=\sum_{i=1}^{n}\alpha_i^j$ and $l=k-1+t$, then we have \\
\begin{eqnarray*}&&
C_{\beta}C_{\beta}^T
= \begin{pmatrix}
k &0& \ldots  & 0 &0\\
0& 0& \ldots   & 0  & \beta^k k \\
\vdots& \vdots& \ldots  & \vdots &\vdots\\
0&\beta^k k&\cdots  & 0&0 \\
\end{pmatrix} \\
&+&
\resizebox{9cm}{!}{$\begin{pmatrix}
0 & 0& \ldots  &0& \eta\beta^l\theta_{l}&0  &\ldots&0\\
0 & 0& \ldots  &0& \eta\beta^{l+1}\theta_{l+1}&0 &\ldots&0\\
\vdots & \vdots& \ldots  &\vdots& \vdots & \vdots& \ldots  &\vdots\\
0 & 0& \ldots  &0& \eta\beta^{l+h-1}\theta_{l+h-1}&0  &\ldots& 0\\
\eta\beta^l\theta_{l}& \eta\beta^{l+1}\theta_{l+1}& \ldots  &\eta\beta^{l+h-1}\theta_{l+h-1}&
2\eta\beta^{l+h}\theta_{l+h}+\eta^2\beta^{2l}\theta_{2l}&\eta\beta^{l+h+1}\theta_{l+h+1}&\ldots&\eta\beta^{l+k-1}\theta_{l+k-1}\\
0 &0& \ldots  &0& \eta\beta^{l+h+1}\theta_{l+h+1}&0 &\ldots&0\\
\vdots & \vdots& \ldots  &\vdots & \vdots & \vdots& \ldots  &\vdots\\
0 & 0& \ldots  &0 & \eta\beta^{l+k-1}\theta_{l+k-1}&0 &\ldots&0
\end{pmatrix}$.}
\end{eqnarray*} Since every $\theta_t$ for $l\leq t\leq l+k-1$ is zero except exactly one $\theta_{t'}$, we
can rewrite 
\begin{eqnarray*}&&C_{\beta}C_{\beta}^T=\left( \begin{array}{cccccc}
k &0& \ldots  & 0 &0\\
0& 0& \ldots   & 0  & \beta^k k \\
\vdots& \vdots& \ldots  & \vdots &\vdots\\
0&\beta^k k&\cdots  & 0&0 \\
\end{array} \right)\\
&+&\left( \begin{array}{cccccccc}
0 &\ldots&0& \ldots  & 0 &\ldots&0\\
\vdots&& \vdots& \ldots  & \vdots &&\vdots\\
0&\ldots &0& \ldots   & *_{\beta}  & \ldots&0 \\
\vdots&& \vdots& \ldots  & \vdots &&\vdots\\
0&\ldots &*_{\beta}& \ldots   & \Delta_{\beta}  &\ldots& 0 \\
\vdots&& \vdots& \ldots  & \vdots &&\vdots\\
0&\ldots&0&\cdots  & 0&\ldots&0 \\
\end{array} \right),\end{eqnarray*}
where $\ast_{\beta}$ and $\Delta_{\beta}$ are all elements in $\Bbb F_q$, the  $\ast_{\beta}$ and $\Delta_{\beta}$ are respectively entries located in the $(i+1,h+1)th$, $(h+1,i+1)th$ and $(h+1,h+1)th$ positions, and the other elements are all zero.

Let $G=[C_1: C_{\gamma}]$ and $h>0$. Then \begin{eqnarray*}&&GG^T=C_1C_1^T+C_{\gamma}C_{\gamma}^T
\\
&=&\left( \begin{array}{cccccc}
2k &0& \ldots  & 0 &0\\
0& 0& \ldots   & 0  & (1+\gamma^k) k \\
\vdots& \vdots& \ldots  & \vdots &\vdots\\
0&(1+\gamma^k) k&\cdots  & 0&0 \\
\end{array} \right)\\
&+&\left( \begin{array}{cccccccc}
0 &\ldots&0& \ldots  & 0 &\ldots&0\\
\vdots&& \vdots& \ldots  & \vdots &&\vdots\\
0&\ldots &0& \ldots   & *_{1} + *_{\gamma}& \ldots&0 \\
\vdots&& \vdots& \ldots  & \vdots &&\vdots\\
0&\ldots &*_{1} + *_{\gamma}& \ldots   &  \Delta_1+\Delta_{\gamma}  &\ldots& 0 \\
\vdots&& \vdots& \ldots  & \vdots &&\vdots\\
0&\ldots&0&\cdots  & 0&\ldots&0 \\
\end{array} \right).\end{eqnarray*}
It is easy to find an elementary matrix $P$ such that \begin{eqnarray*}&&PGG^TP^T =
\left( \begin{array}{cccccc}
2k &0& \ldots  & 0 &0\\
0& 0& \ldots   & 0  & (1+\gamma^k) k \\
\vdots& \vdots& \ldots  & \vdots &\vdots\\
0&(1+\gamma^k) k&\cdots  & 0&0 \\
\end{array} \right)\\
&+&\left( \begin{array}{cccccccc}
0 &\ldots&0& \ldots  & 0 &\ldots&0\\
\vdots&& \vdots& \ldots  & \vdots &&\vdots\\
0&\ldots &0& \ldots   & 0& \ldots&0 \\
\vdots&& \vdots& \ldots  & \vdots &&\vdots\\
0&\ldots &0& \ldots   &  \Delta_1+\Delta_{\gamma}  &\ldots& 0 \\
\vdots&& \vdots& \ldots  & \vdots &&\vdots\\
0&\ldots&0&\cdots  & 0&\ldots&0 \\
\end{array} \right).\end{eqnarray*}

By the given conditions,  we have $k\mid (q-1)$, $k< (q-1)$, $q$ is even, and $\gamma $ is a primitive element of $\Bbb F_q$. Hence, $\gamma ^k+1\neq0$ and $2k=0$.  Then we have  $\mbox{rank}(GG^T)=k-1$.  The result follows from Lemma 2.1.
$\blacksquare$


\begin{lemma}\label{lem7}{\rm  Let $\Bbb F_q$ be a finite field of odd order $q$ and $k$ be  a positive integer with $k\mid (q-1)$ and $2<k<{(q-1)}/{2}$. If  $h>1$, then there exists a $[2k,k-1]_q$  twisted Reed-Solomon code $\mathcal C_{k-1}(\boldsymbol{\alpha}, t,h,\eta)$ over $\Bbb F_q$ with one-dimensional hull   for
$\boldsymbol{\alpha}=(\alpha_1,\ldots,\alpha_k,\gamma\alpha_1,\ldots,\gamma\alpha_k)$,
where $\gamma$ is a primitive element of $\Bbb F_q$ and $\alpha_i=\gamma^{\frac{q-1}{k}i}$ for $1\le i\le k$.

}

\end{lemma} 

{\bf Proof} Let \begin{equation}\resizebox{8cm}{!}{$D_{\beta}=\left( \begin{array}{cccccc}
1 & 1& \ldots  & 1 &1\\
\beta\alpha_1& \beta\alpha_2& \ldots   &  \beta\alpha_{k-1}  & \beta\alpha_k \\
\vdots& \vdots& \ldots  & \vdots &\vdots\\
(\beta\alpha_1)^{k-2}& (\beta\alpha_2)^{k-2} &\cdots  & (\beta\alpha_{k-1})^{k-2} &(\beta\alpha_k)^{k-2} \\
\end{array} \right)$.}\end{equation} Then \begin{eqnarray*}&&
D_{\beta}D_{\beta}^T
= \begin{pmatrix}
k &0 &0& \ldots  & 0 &0\\
0 &0 &0& \ldots  & 0 &0\\
0 &0& 0& \ldots   & 0  & \beta^k k \\
\vdots& \vdots& \ldots  & \vdots &\vdots\\
0 &0&\beta^k k&\cdots  & 0&0 \\
\end{pmatrix}. \end{eqnarray*}
Let $H_{\beta}=D_{\beta}+E_{\beta}$, where
\begin{eqnarray*}\resizebox{8.9cm}{!}{$
E_{\beta}=\left( \begin{array}{cccccc}
0 & 0& \ldots  & 0 &0\\
\vdots& \vdots& \ldots  & \vdots &\vdots\\
\eta(\beta\alpha_1)^{k-2+t}& \eta(\beta\alpha_2)^{k-2+t}& \ldots   &  \eta(\beta\alpha_{k-1})^{k-2+t}  &\eta(\beta\alpha_k)^{k-2+t} \\
\vdots& \vdots& \ldots  & \vdots &\vdots\\
0 & 0& \ldots  & 0 &0\\
\end{array} \right)\begin{array}{l} \\   \\\leftarrow(h+1)th\\  \\ \\  \end{array}$}.\end{eqnarray*}

 Let $G=[H_1:H_{\gamma}]$. By the proof of Lemma 3.1,  \begin{eqnarray*}&&GG^T=H_1H_1^T+H_{\gamma}H_{\gamma}^T
\\
&=&\left( \begin{array}{cccccccc}
2k &0 &0& \ldots  & 0 &0&0\\
0 &0 &0& \ldots  & 0 &0&0\\
0 &0& 0& \ldots   & 0  & (1+\gamma^k) k &0\\
\vdots& \vdots& \ldots  & \vdots &\vdots\\
0 &0&(1+\gamma^k) k&\cdots  & 0&0&0 \\
\end{array} \right)\\
&+&\left( \begin{array}{cccccccc}
0 &\ldots&0& \ldots  & 0 &\ldots&0\\
\vdots&& \vdots& \ldots  & \vdots &&\vdots\\
0&\ldots &0& \ldots   & *_{1} + *_{\gamma}& \ldots&0 \\
\vdots&& \vdots& \ldots  & \vdots &&\vdots\\
0&\ldots &*_{1} + *_{\gamma}& \ldots   &  \Delta_1+\Delta_{\gamma}  &\ldots& 0 \\
\vdots&& \vdots& \ldots  & \vdots &&\vdots\\
0&\ldots&0&\cdots  & 0&\ldots&0 \\
\end{array} \right).\end{eqnarray*} Note that $q$ is odd and $k<\frac{q-1}{2}$. We have $2k\neq 0$ and $1+\gamma^k\neq0$. By the same process of the proof of Lemma 3.1, the rank of $GG^T$ is $k-1$ and result follows from Lemma 2.1.
$\blacksquare$

\begin{remark} By the process of Lemma 3.2, we can also construct some twisted Reed-Solomon codes with small hulls.
\end{remark}

An effective method for construction of twisted Reed-Solomon codes with MDS property
is to use the lifting  of the finite field (refer to \cite{BPR}).   Hence, we obtain the following theorem by Lemmas 2.5, 2.7, 3.1 and 3.2.

\begin{theorem} {\rm Let $q$ be a power of a prime and $\Bbb F_s \subset \Bbb F_q$ with
$|\Bbb F_q \backslash \Bbb F_s|>6$. Suppose that  $k$ is a positive integer with $k\mid (q-1)$. 


(1) If $q$ is even and  $2<k< (s-1)$,
then there exists a $[2k, k]_q$  MDS non-Reed-Solomon code with one-dimensional hull. 

(2) If $q$ is odd and  $2<k< (s-1)/2$,
then there exists a $[2k, k-1]_q$  MDS non-Reed-Solomon code with one-dimensional hull.

}
\end{theorem}

%
%


In the following, we will present some examples to show our main results.
\begin{example}
{\rm
(1) Let $q=2^4=16$, $k=5$, and $\gamma$ be a primitive element of $\Bbb F_{q}$. Consider a twisted Reed-Solomon code $\mathcal C_5(\boldsymbol{\alpha}, 1,3,\eta)$, when $\boldsymbol{\alpha}=(1,\gamma^{3}, \gamma^{6}, \gamma^{9},  \gamma^{12},\gamma,\gamma\gamma^{3}, \gamma\gamma^{6}, \gamma\gamma^{9},\gamma\gamma^{12})$ and $\eta=\gamma^i\in \Bbb F_{16}$. 
By Lemma 3.1, $\mathcal C_5(\boldsymbol{\alpha}, 1,3,\gamma^i)$ has one-dimensional hull for all $i$. By Magma, it follows that
the codes $\mathcal C_5(\boldsymbol{\alpha}, 1,3,\eta)$ are not MDS codes, which have parameters $[10,5,5]_{16}$.

(2) Let $q=2^8=256$, $k=5$, and $w$ be a primitive element of $\Bbb F_{q}$ and $\gamma=w^{17}\in \Bbb F_{16}$. Consider a twisted Reed-Solomon code $\mathcal C_5(\boldsymbol{\alpha}, 1,3,\eta)$, when $\boldsymbol{\alpha}=(1,\gamma^{3}, \gamma^{6}, \gamma^{9},  \gamma^{12},\gamma,\gamma\gamma^{3}, \gamma\gamma^{6}, \gamma\gamma^{9},\gamma\gamma^{12})$ and $\eta=w^i\in \Bbb F_{256}$. 
By Lemma 3.1, $\mathcal C_5(\boldsymbol{\alpha}, 1,3,w^i)$ has one-dimensional hull for all $i$. By Magma and Lemmas 2.5 and 2.7, there exists an integer $i$ with $17\nmid i$ such that 
the code $\mathcal C_5(\boldsymbol{\alpha}, 1,3,\eta)$ is an   MDS  non-Reed-Solomon code with parameters $[10,5,6]_{256}$.

}

\end{example}

\begin{example}
{\rm
(1) Let $q=3^4=81$, $k=5$, and $\gamma$ be a primitive element of $\Bbb F_{q}$. Consider a twisted Reed-Solomon code $\mathcal C_4(\boldsymbol{\alpha}, 2,2,\eta)$, when $\boldsymbol{\alpha}=(1,w, w^{2}, w^{3},  w^{4},\gamma,\gamma w, \gamma w^{2}, \gamma w^{3},\gamma w^{4})$ and $\eta=\gamma^i\in \Bbb F_{81}$ and $w=\gamma^{16}$. 
By Lemma 3.2, $\mathcal C_4(\boldsymbol{\alpha}, 2,2,\gamma^i)$ has one-dimensional hull for all $i$. By Magma, it follows that
the codes $\mathcal C_4(\boldsymbol{\alpha}, 2,2,\gamma^i)$ are MDS with parameters $[10,4]_{81}$ when $\eta$ belongs to $H=\{\gamma^j: j=0,6,16,22,32,38,48,54,64,70\}$.
  Since $|H|>6$, by Lemma 2.6, there exists $\eta\in H$ such that $\mathcal C_4(\boldsymbol{\alpha}, 2,2,\eta)$ is a non-Reed-Solomon code. As a result, there exists a $[10,4]_{81}$   MDS non-Reed-Solomon code with one-dimensional hull.

(2) Let $q=3^8=6561$, $k=5$, and $\theta$ be a primitive element of $\Bbb F_{q}$, $\gamma=\theta^{82}\in \Bbb F_{81}$, and $w=\gamma^{16}$. Consider a twisted Reed-Solomon code $\mathcal C_4(\boldsymbol{\alpha}, 2,2,\eta)$, when $\boldsymbol{\alpha}=(1,w, w^{2}, w^{3},  w^{4},\gamma,\gamma w, \gamma w^{2}, \gamma w^{3},\gamma w^{4})$ and $\eta=\theta^i\in \Bbb F_{6561}$. 
By Lemma 3.2, $\mathcal C_4(\boldsymbol{\alpha}, 2,2,\gamma^i)$ has one-dimensional hull for all $i$. By Magma and Lemmas 2.5 and 2.7, there exists an integer $i$ with $82\nmid i$ such that 
the code $\mathcal C_4(\boldsymbol{\alpha}, 2,2,\gamma^i)$ is an  MDS  non-Reed-Solomon code with parameters $[10,4,7]_{6561}$.

}

\end{example}

\section{Concluding remarks}

For a given linear code, in general case it is hard to show that  if the code is monomially equivalent to a Reed-Solomon code with the same parameters.
In this paper, we applied twisted Reed-Solomon codes to construct  some MDS codes, which  have one-dimensional hull and are not monomially equivalent to Reed-Solomon codes. We also presented some examples by using Magma.





%

\section*{Acknowledgment}


The author is very grateful to the reviewers and the Associate Editor for their valuable comments and suggestions to improve the quality of this paper.

\ifCLASSOPTIONcaptionsoff
  \newpage
\fi

\end{document}